# Argon difluoride (ArF$_2$) stabilized at high pressure

Dominik Kurzydłowski[1,2*], Patryk Zaleski-Ejgierd[3*]

[1]Centre of New Technologies, University of Warsaw, ul. S. Banacha 2c, 02-097, Warsaw, Poland

[2] Faculty of Mathematics and Natural Sciences, Cardinal Stefan Wyszynski University in Warsaw, ul. K. Wóycickiego 1/3, 01-938, Warsaw, Poland

[3]Institute of Physical Chemistry, Polish Academy of Sciences, ul. M. Kasprzaka 44/52 01-224, Warsaw, Poland

E-mail: d.kurzydlowski@cent.uw.edu.pl, pze.work@gmail.com

*On account of the rapid development of noble gas chemistry in the past half-century both xenon and krypton compounds can now be isolated in macroscopic quantities. The same though does not hold true for the next lighter group 18 element, argon, which forms only isolated molecules stable solely in low-temperature matrices or supersonic jet streams. With the use of the HSE06 hybrid DFT method we show that argon difluoride can be stabilized at high pressure from an equimolar mixture of Ar and F$_2$. The compound contains nearly linear F-Ar-F moieities analogous in structure and electronic properties to KrF$_2$ and XeF$_2$.*

## 1. Structure of ArF$_2$ at high pressure

We have assumed two possible structure of compressed ArF$_2$ in the pressure range studied (0 – 100 GPa). One of them is isostructural with the ambient pressure structure of XeF$_2$ (*I4/mmm* symmetry, Fig. 1a-c), the other is analogous to the high pressure polymorph of XeF$_2$ of *Pnma* symmetry (Fig. 1d-e) which is predicted to be more stable than *I4/mmm* above 105 GPa [1]. In our calculations the *Pnma* structure converges to *I4/mmm* at 0 GPa, but preserves a distinct geometry at higher pressure. The tetragonal *I4/mmm* polymorph (characterized by cell vectors $\boldsymbol{a_T} = \boldsymbol{b_T} \ll \boldsymbol{c_T}$, Fig. 1a) can be described as built from layers extending in $\boldsymbol{a_T}$ and $\boldsymbol{b_T}$ directions in which linear and symmetric ArF$_2$ molecules are placed on a square lattice, with the Ar-F bonds perpendicular to the layer (Fig. 1c). Along $\boldsymbol{c_T}$ neighbouring layers are shifted with respect to each other by ($\boldsymbol{a_T}$ + $\boldsymbol{b}_T$)/2.

Transforming the *I4/mmm* polymorph to a √2x√2x1 cell (with ***a*** = ***a***$_T$ + ***b***$_T$, ***b*** = ***a***$_T$ − ***b***$_T$ and ***c*** = ***c***$_T$) enables comparison with the orthorhombic *Pnma* structure (Fig. 1b, d). The latter also consists of stacked ArF$_2$ layers, but the molecules are slightly bent (179° − 175° from 5 to 100 GPa). Moreover movement of half of them along the ***a*** direction results in considerable shortening of two intra-layer Ar⋯Ar contacts at the expense of the other two (Fig. 1e). As a result of this distortion the ***a*** cell vector of *Pnma* expands while ***b*** shortens compared to ***a*** in the √2x√2x1 cell of *I4/mmm* (Fig. 2). Together with a small contraction along ***c*** this makes the volume of *Pnma* approximately 1.5 % smaller than *I4/mmm* in the whole pressure range studied.

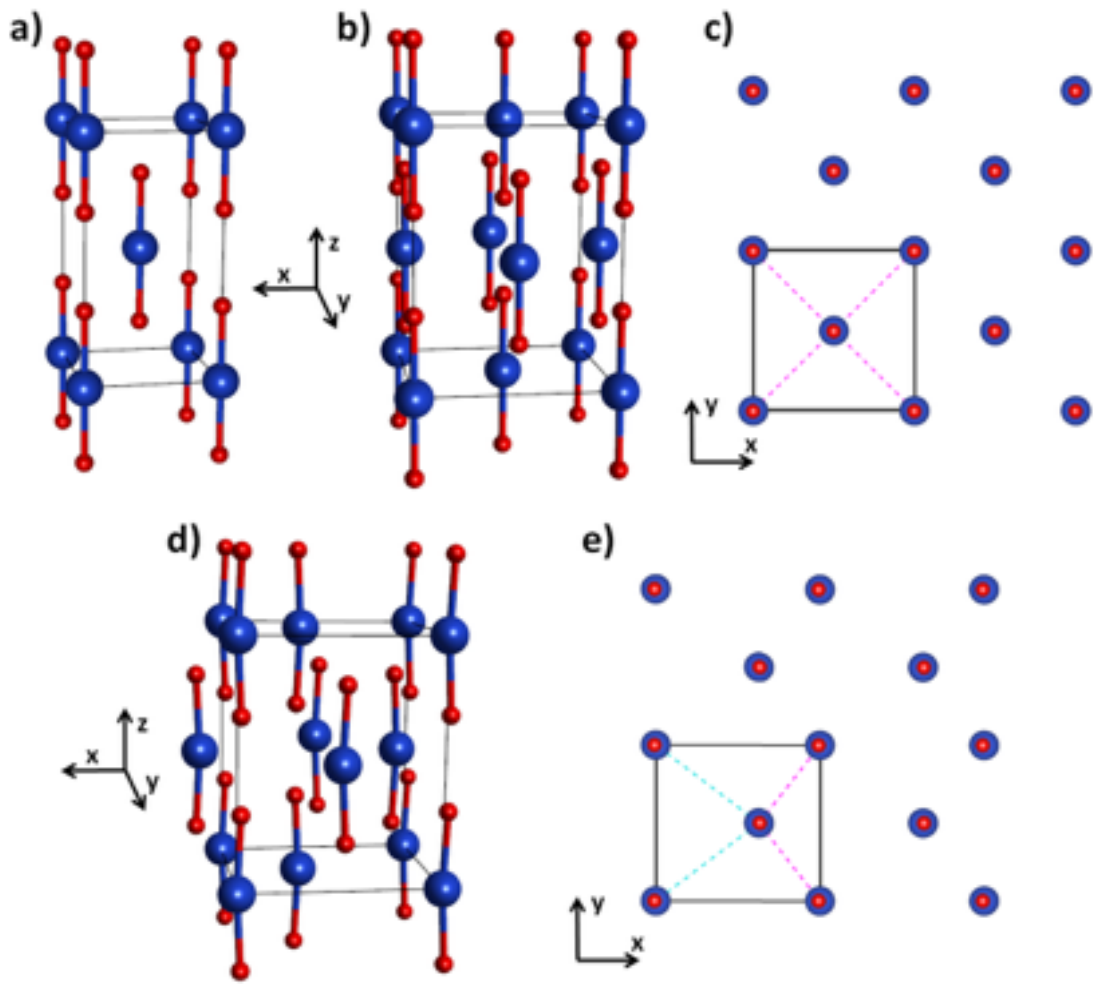

**Fig. 1 Perspective view of *I4/mmm* in the conventional (a) and √2x√2x1 cell (b) together with a view of this supercell along the *c* axis (c). The primitive cell of *Pnma* (d) and its view along *c* (e) are also shown. Blue/red balls mark Ar/F atoms; Ar⋯Ar contacts are marked with dashed lines.**

It's noteworthy to point that the symmetry of *Pnma* does not restrict both Ar-F bonds of the ArF$_2$ molecule to be equal. In fact it is predicted that for the *Pnma* structure of XeF$_2$ alternation of the Xe-F bond lengths leads to formation of an [Xe–F]$^+$F$^-$ ionic compound at 200 GPa [1]. In case of ArF$_2$ no signs of such a process are observed – up to 100 GPa the Ar-F bonds in *Pnma* are equal to each and differ by no more than 0.1 % from the ones of *I4/mmm* (see Fig. S 1 in Supplementary Information, SI).

The length of the Ar-F bond in solid ArF$_2$ at 0 GPa is 1.764 Å, compared with 1.761 Å calculated for the free molecule. Both values are 5 % smaller than the calculated bond length of isoelectronic ClF$_2^-$ (1.857 Å) [2], which is not surprising given the fact that ArF$_2$ is a neutral specie (for the BrF$_2^-$/KrF$_2$ pair, the analogous difference is 4 % [3]). Upon compression to 100 GPa the Ar-F bond length decreases by 6 %, to 1.663 Å (Fig. S 1).

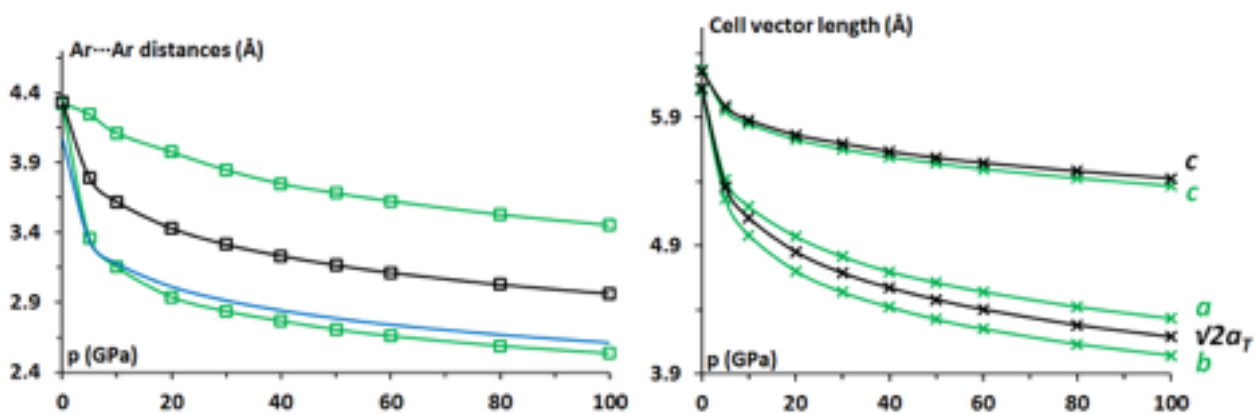

**Fig. 2 Pressure dependence of intra-layer Ar⋯Ar distances (left) and cell vector lengths (right) for *I4/mmm* (black lines) and *Pnma* (green lines). The blue line depicts the calculated pressure dependence of the closest Ar⋯Ar separation in solid Ar.**

Surprisingly we were able to find only one theoretical prediction of the bond length and dissociation energy of ArF$_2$ in the work on noble gas difluorides by Euler *et al*. These authors obtain an Ar-F bond length of 1.857 Å and a dissociation energy of 2.95 eV. In contrast, from our calculations the energy of the ArF$_2$ molecule lies 0.1 eV above that of the Ar + 2F˙ system. Together with the dissociation energy of F$_2$ (Table 1) this translates to an energy of formation ($E_f$) of 1.5 eV[4].

## 2. High pressure synthesis

The enthalpy of formation ($H_f$) of ArF$_2$ at a given pressure has been calculated as the difference in enthalpy of *I4/mmm* and *Pnma* with respect to solid F$_2$ and Ar.

As can be seen in Fig. 3 the *I4/mmm* and *Pnma* polymorphs of ArF$_2$ become more stable than Ar + F$_2$ at 71 and 57 GPa, respectively. *Pnma* has a lower enthalpy than the tetragonal phase by 9 meV per fu at 5 GPa up to 254 meV at 100 GPa. This is by far the effect of the smaller volume of *Pnma* – at 100 GPa the difference in $E_f$ between the two structures (73 meV per fu) constitutes only 29 % of the difference in $H_f$ with the remaining part coming from differences in the $pV$ term. Thus one can conclude that for the orthorhombic structure the small departure of the F-Ar-F angle from 180° does not have a major impact on the energy of the ArF$_2$ units.

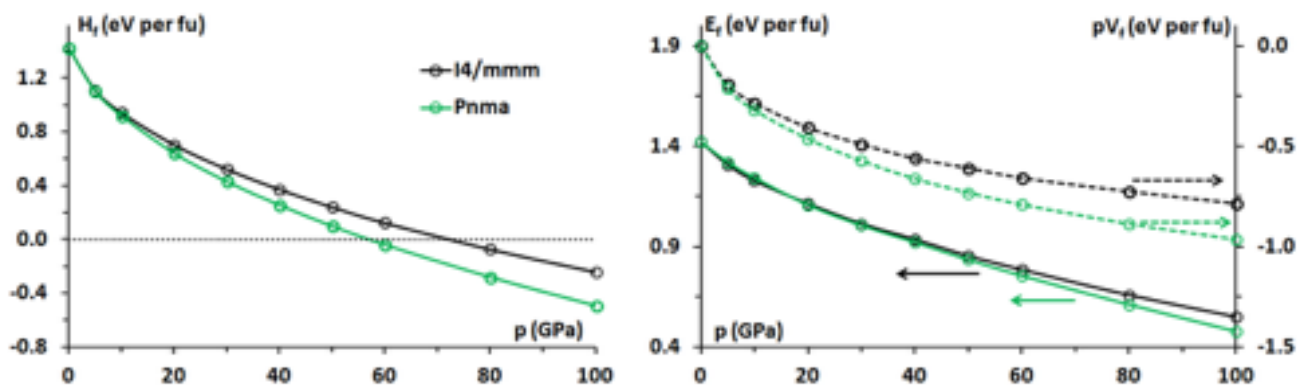

**Fig. 3 Enthalpy of formation (left) together with the energy of formation and the pV$_f$ term (right) of *I4/mmm* and *Pnma* polymorphs of ArF$_2$ (black, green lines respectively).**

The difference in volumes between *I4/mmm*/*Pnma* and the Ar + F$_2$ system (Fig. 4), is one of the main driving forces for the synthesis of ArF$_2$. At 60 GPa the formation of the *Pnma* polymorph is accompanied by a volume reduction of 7 %. As can be seen in Fig. 4 this reaction should be clearly distinguishable by synchrotron powder x-ray diffraction, mainly by the disappearance of strong reflexes originating from solid Ar.

It's noteworthy to point that pressures in the range of 60 – 70 GPa are now easily accessible by standard experimental techniques utilizing diamond anvil cells (DACs). Despite the very reactive and corrosive nature of fluorine this compound has been compressed in a DAC in the past [5]. In case of the ArF$_2$ synthesis a much less reactive Ar/F$_2$ mixture (analogous to N$_2$/F$_2$ mixtures used in commercial processing of plastics [6]) would be loaded into a DAC. Furthermore high pressure experiments conducted on XeF$_2$ [7] show that by choosing an appropriate gasket one can minimalize side reaction during compression of strong oxidizer . Finally, we note that the proximity of the boiling points of Ar and F$_2$ (87.3 and 85.0 K, respectively) should enable efficient lowtemperature loading of an Ar/F$_2$ mixture into a DAC.

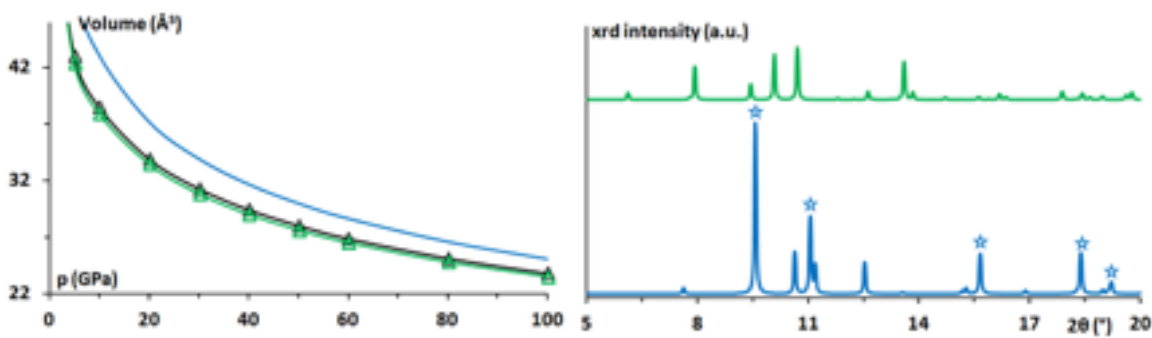

**Fig. 4** Left: Pressure dependence of volume of Ar + F$_2$ (blue line) and *I4/mmm*/*Pnma* (black/green line). Right: comparison of the diffraction pattern at 60 GPa of an equimolar mixture of *fcc* Ar and *Cmca* F$_2$ (blue line, stars mark Ar reflexes), and *Pnma* (green line).

### 3. Electronic structure

According to HSE calculations *I4/mmm* is a wide-gap insulator at 0 GPa with a band gap of 5.6 eV[8]. As can be seen in the electronic density of states (DOS, Fig. 5) the states below −17 eV with respect to the Fermi have *s* character, while does lying at high energies have *p* character. Analysis of the contributions from $p_x$, $p_y$ and $p_z$ orbitals centred on different atoms clearly shows that the character of electronic states of ArF$_2$ molecules can be understood within the framework of the electron-rich three-centre model [9]. As shown in Fig. 5 one can identify occupied states as originating from bonding ($\sigma$, $\pi$), non-bonding ($\sigma_n$, $\pi_n$), and anti-bonding ($\sigma^*$, $\pi^*$) combinations of both *s* and *p* atomic orbitals. In agreement with the three-centre model the empty conduction band is derived from the $\sigma^*$ combination of $p_z$ orbitals (Fig. 6).

Compression of *I4/mmm* to 100 GPa does not have a major impact on the band gap which decreases only slightly to 5.4 eV[†]. Also the nature of the electronic states remains largely unchanged with the separation between *s* and *p* states still clearly seen (Fig. 6). The much larger compression of the $a_T$ lattice parameter compared to $c_T$ (−32% and −13 %, respectively, see Fig. 2) results in bigger broadening of the $p_x$ and $p_y$ states compared to $p_z$ (Fig. 6).

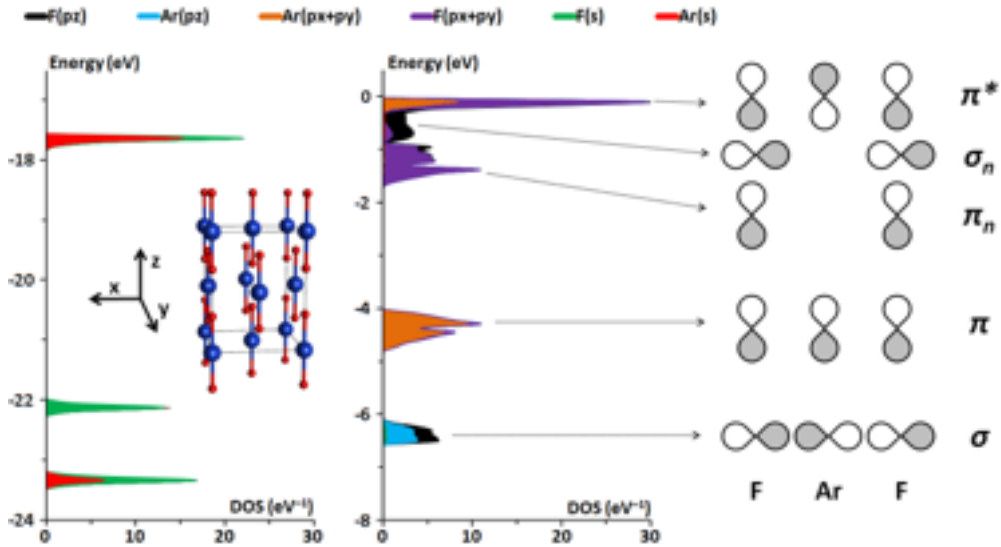

**Fig. 5** Total and partial electronic DOS of *I4/mmm* at 0 GPa (left and middle) together with a schematic depiction of the $\sigma(p)$ and $\pi(p)$ orbitals of the ArF$_2$ molecule (right).

The electronic structure of *Pnma* at 100 GPa is similar to that of *I4/mmm* (Fig. 6). The most notable differences are that the orthorhombic structure exhibits a band gap 0.8 eV smaller (4.6 eV) and some mixing of *s* and $p_z$ states in bands located in the –2 to –4 eV and –24 to –26 eV energy intervals.

The electronic structure of ArF$_2$ can be compared with that of XeF$_2$ – another three-centre electron-rich system. Due to the large differences in energies of fluorine and xenon *s* orbitals the latter system does not exhibit any hybridization between *s* states, even at high pressure [1]. Moreover large differences in Xe and F electronegativity ($\Delta\chi_{F,Xe}$ = 1.18 in the Mulliken scale [10]) result in substantial ionic contribution to the Xe-F bond. The Ar/F pair is characterized by $\Delta\chi_{F,Ar}$ of 0.71 and thus Ar-F bonds are much more covalent. This explains why, in contrast to XeF$_2$ [1], ArF$_2$ is much less susceptible to 'self-dissociation' into [ArF]$^+$F$^-$ at high pressure. Finally we note that ArF$_2$ remains an insulator up to 100 GPa, while for XeF$_2$ calculations indicate a band gap closure at 50 GPa [1].

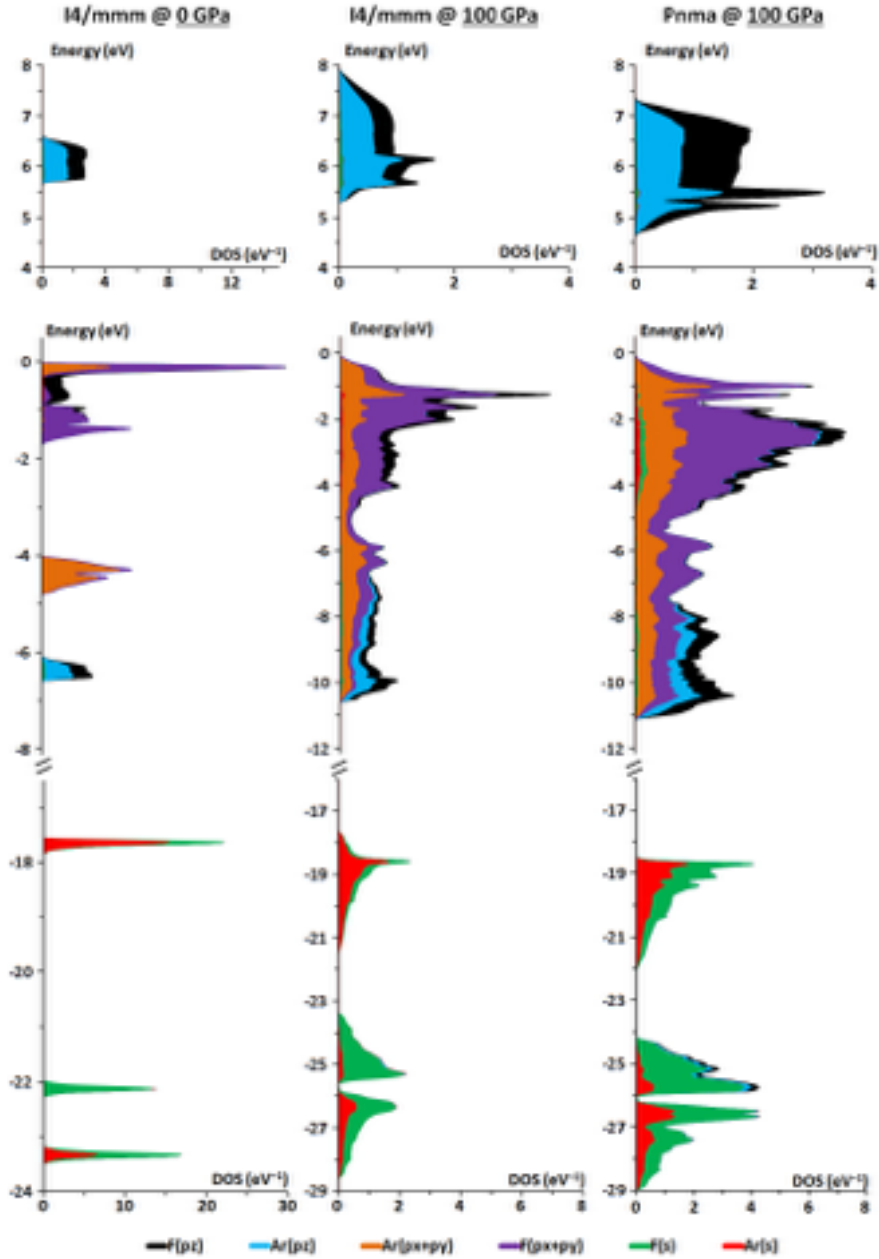

**Fig. 6** Electronic DOS of for *I4/mmm* (0, 100 GPa) and *Pnma* (100 GPa).

## 4. Calculation details

Periodic DFT calculations utilized the HSE06 hybrid potential, while the PBE exchangecorrelation functional was used for comparative calculations. The projectoraugmented-wave (PAW) method [11] was used, as implemented in the *VASP 5.2* code.[12] The cut-off energy of the plane waves was set to 1200 eV with a selfconsistent-field convergence criterion of $10^{-6}$ eV. Valence electrons were treated explicitly, while VASP pseudopotentials were used for the description of core electrons. In order to ensure proper treatment of valence electrons a 'hard' pseudopotential for fluorine (PAW radial cut-off of 1.1 Å) has been chosen. The *k*-point mesh was set at *2π* x 0.06 Å⁻¹, and was increased to *2π* x 0.04 Å⁻¹ for DOS calculations. All structures were optimized using a conjugategradient algorithm until the forces acting on the atoms were smaller than 5 meV/Å. The abovementioned parameters ensured convergence of the calculated enthalpy within 2 meV per atom. The enthalpy of formation of ArF$_2$ was calculated with respect to elemental Ar in either the *fcc* or *hcp* structure [13], as well as the α polymorph of F$_2$ [14]. We note that above 50 GPa α-F$_2$ (*C2/c* space group) symmetrizes spontaneously to a *Cmca* structure which is analogous to the high-pressure polymorph of Cl$_2$ [15].   Calculations of free molecules were conducted by placing a given molecule in the centre of a cell large enough to allow for minimum 10 Å separation between adjacent molecules.

The importance of using the HSE functional in calculations of thermodynamic stability of fluorine bearing compounds can be easily seen by comparing the dissociation energy of a free F$_2$ molecule calculated with the PBE and HSE functionals (Table 1). The former severely overestimates the dissociation energy, while the latter gives a value much closer to the experimental one, and differing only slightly from that obtained in molecular B3LYP calculations.

|  | Exp | PBE | HSE | B3LYP [] |
|---|---|---|---|---|
| **D$_e$ (eV)** | 1.606 | 2.766 | 1.444 | 1.578 |
| **R$_e$ (Å)** | 1.412 | 1.414 (0.1 %) | 1.376 (–2.5 %) | 1.396 (–1.1 %) |

**Table 1 Comparison of dissociation energy (without ZPE correction) and bond length obtained for the F$_2$ molecule with different functionals.**

The F-F bond length obtained with the PBE functional seems to be in fortuitous agreement with the experimental value given the fact that GGA methods tend to overestimate bond lengths. Again the HSE value is close to the B3LYP one. Interestingly both slightly underestimate the F-F bond length.

Geometry optimization of the molecular crystal of fluorine (in the ordered α phase [16]) again shows that the HSE values are in better agreement with experiment – for example the cell volume is overestimated by 2 % compared to 16 % overestimation with the PBE functional.

## 5. Supplementary information

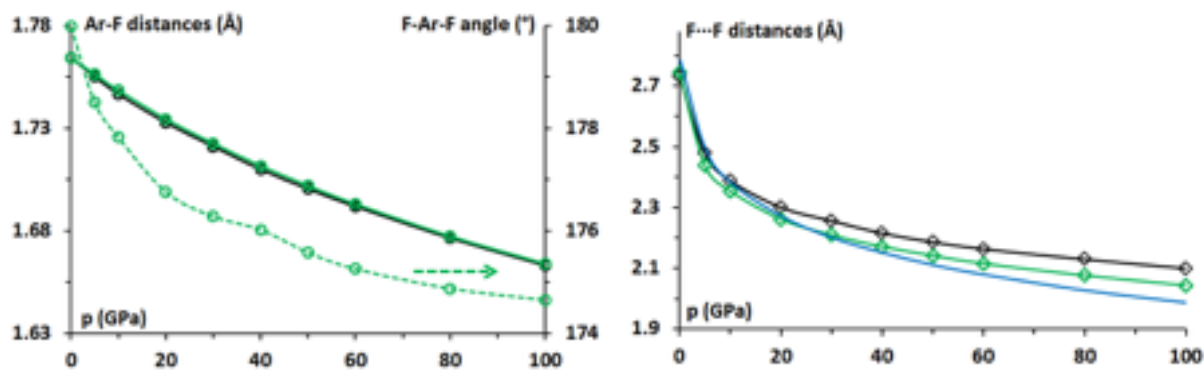

**Fig. S 1 Left: Ar-F bond length and F-Ar-F angle; right: shortest F⋯F distances. Black/green lines correspond to *I4/mmm*/*Pnma*, blue line depicts the calculated pressure dependence of the F⋯F contacts in solid $F_2$.**

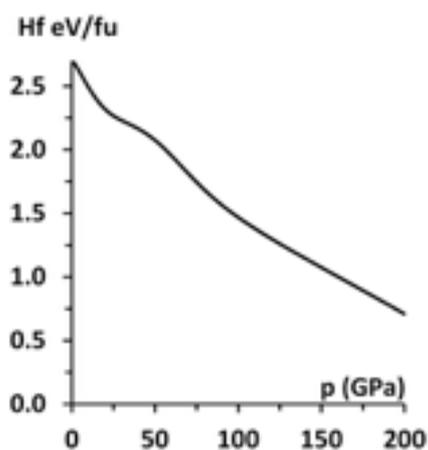

The figure above shows the pressure dependence of the enthalpy of formation of $ArF_2$ from a mixture of $NF_3$ and Ar (*i.e.* the enthalpy change associated with the reaction: $Ar + ⅔NF_3 \rightarrow ArF_2 + ⅓N_2$) calculated at the HSE06 level. As can be seen, the formation of $ArF_2$ and $N_2$ from the mixture of $NF_3$ and Ar is disfavored within the whole studied pressure range (0 − 200 GPa). A straightforward extrapolation to higher pressures indicates that this reaction should proceed spontaneously only above 300 GPa. In the calculations $NF_3$ has been assumed to adopt a recently proposed high-pressure structure [17]. For $N_2$ below 60 GPa a $P4_12_12_1$ molecular structure [18] has been assumed while above that pressure the polymeric cubicgauche polymorph [19] was taken as the ground state.